\documentclass[12pt]{article}
\usepackage{amsmath}
\usepackage{graphicx}
\usepackage{geometry}
\usepackage[applemac]{inputenc}

\usepackage{natbib}

\ifx\pdfoutput\@undefined\usepackage[usenames,dvips]{color}
\else\usepackage[usenames,dvipsnames]{color}
\IfFileExists{pdfcolmk.sty}{\usepackage{pdfcolmk}}{} 
\fi
\usepackage[plainpages=false,pdfpagelabels,pagebackref=false,naturalnames=true,hyperindex=true]{hyperref}
\hypersetup{colorlinks=true,
urlcolor=Cerulean,linkcolor=BrickRed,citecolor=RoyalBlue,
  pdfpagemode=None,
  pdfstartview=FitH}
\usepackage[all]{hypcap}

\begin{document}

\title{The Past, Present and Future\\ of Cybernetics and Systems Research}
\author{Carlos Gershenson$^{1}$, P\'eter Csermely$^{2}$, P\'eter \'Erdi$^{3,4}$,\\
 Helena Knyazeva$^{5}$, and Alexander Laszlo$^{6}$  \\
$^{1}$Instituto de Investigaciones en Matem\'aticas Aplicadas y en Sistemas \\
\& Centro de Ciencias de la Complejidad, \\
Universidad Nacional Aut\'onoma de M\'exico.\\
$^{2}$ Semmelweis University, Department of Medical Chemistry, Budapest, Hungary. \\
$^{3}$ Center for Complex Systems Studies, Kalamazoo College, Kalamazoo, MI, USA.\\
$^{4}$ Institute for Particle and Nuclear Physics, Wigner Research Centre for Physics,\\ 
Hungarian Academy of Sciences, Budapest, Hungary.\\
$^{5}$Institute of Philosophy, Russian Academy of Sciences, Moscow, Russia.\\
$^{6}$Syntony Quest, Sebastopol, CA, USA.
}
\maketitle

\begin{abstract}
Cybernetics and Systems Research (CSR) were developed in the mid-twentieth century, offering the possibility of describing and comparing different phenomena using the same language. The concepts which originated in CSR have spread to practically all disciplines, many now used within the scientific study of complex systems. CSR has the potential to contribute to the solution of relevant problems, but the path towards this goal is not straightforward. 

This paper summarizes the ideas presented by the authors during a round table in 2012 on the past, present and future of CSR.
\end{abstract}

\section{Introduction}

The ideas contained in this paper were presented at a round table with the same title on April $12^{th}$, 2012, during the European Meeting on Cybernetics and Systems Research at the University of Vienna, Austria.
The guided reflection on the challenges and opportunities of cybernetics and systems research (CSR) included initial interventions by panelists Peter Erdi, Helena Knyazeva, Stefan Thurner, Peter Csermely, and Alexander Laszlo. Afterwards, the floor was opened to interventions from the general public and further interventions by panelists.

Science strives for understanding our world. This is also the aim of CSR~\citep{HeylighenJoslyn2001}. One of the main differences between traditional science and CSR is that the former focusses more on the structure, while the latter focusses more on processes and dynamics. In this way, the same description can be used to describe different phenomena. In other words, CSR searches for \emph{isomorphisms} across disciplines.

\section{The Past}

The scientific study of systems began with Ludwig von Bertalanffy's General Systems Theory perhaps as early as the 1920's, but became popular after the 1940's \citep{von-Bertalanffy:1968}. By describing the general properties of phenomena independently of their substrate, the same language can be used to describe phenomena from different domains, allowing the search of commonalities, for example between logic circuits and neural networks or between human language and DNA transcription. Moreover, systems research allowed the development of synthetic methods \citep{Steels1993} to complement analytic ones. In an analytic method, a model is abstracted from observations. Then the model is used to make predictions, which are contrasted (verified) with further observations. In a synthetic method, a model is also abstracted from observations. However, this model is used to \emph{build} a system which to be verified should reproduce the observations. Synthetic methods not only provide a further approach for understanding phenomena. They also enable to \emph{engineer} systems which exhibit properties of the studied system.

Cybernetics, as defined by Wiener~\citeyearpar{Wiener1948}, is concerned with the scientific study of control and communication in animals and machines. The term comes from the Greek \emph{kibernetes}, which means steersman. This analogy illustrates one of the main concerns of cybernetics: how can systems be guided in their dynamic environment?~\citep{GershensonDCSOS,Prokopenko:2009,Ay2012Guided-self-org}

The roots of cybernetics can be traced to ancient times. There are examples of artifacts which used the principles later formalized by cybernetics that were built in ancient China, India, and Greece. Thales of Miletus already proposed a holistic worldview, which is also present in oriental philosophies. The ideas exposed in Plato's Republic aimed at proposing how a city state could govern itself. It was in a similar context that Amp\`ere wrote about \emph{cybern\'etique} in 1834, concerned with the study of government and bureaucracies. The concept of feedback had been used in several contexts by the XIXth century: Watt used it for steam engines, Wallace for evolution, Maxwell for physics, and von Uexk\"ull for ethology. In the XXth century, developments in electricity, electronics, control, physics, logic, medicine, physiology, neuroscience, and evolutionary theory, among others, generated the necessity of new organization principles to solve particular problems in each area. However, following general systems theory, many of these problems were very similar once their particular substrate was neglected. In the 1920's, the Russian scholar Alexander A. Bogdanov proposed fascinating ideas which are close to GST in his opus “Tectology”. The term “tectology” coined by him comes from Greek and literally signifies “science of organization”. In his opinion, tectology is aimed to reveal some universal principles of organizational forms, whether forms of life, human behavior and health, languages or economics~\citep{Knyazeva2011The-Russian-Cos}.

Several useful concepts which were studied, developed or formalized within cybernetics are now commonly used in science and even common language: information~\citep{Shannon1948}, open and closed systems, variety~\citep{Ashby1956}, homeostasis~\citep{Bernard1859,Cannon:1932,Ashby1947,Ashby:1960}, self-organization~\citep{Ashby1947sos,Ashby1962}, autopoiesis~\citep{VarelaEtAl1974,Maturana1980,luhmann1986autopoiesis}, synergetics~\citep{Haken1988}, dissipative structures~\citep{NicolisPrigogine1977}, organizations~\citep{Beer1966}, game theory~\citep{vonNeumannMorgenstern1944}, cellular automata~\citep{vonNeumann1966},
isomorphisms~\citep{Macrae1951Cybernetics-and}, experimental epistemology~\citep{mcculloch1965embodiments}, and
computational psychiatry~\citep{Montague2012}.

\section{The Present}

The concepts developed within cybernetics have spread memetically to all sciences, humanities, and beyond.  Everybody speaks of systems, although not necessarily citing von Bertalanffy. Many cybernetic concepts are used but not being named as cybernetics. They have been absorbed by our present worldview.

A case of this can be seen with the scientific study of complexity~\citep{Bar-Yam1997,Mitchell:2009}. It takes an approach similar to cybernetics and systems research~\citep[p. 35--45]{Erdi2008complexity}, but in many cases it does not refer to its roots in CSR.

Complexity comes from the Latin \emph{plexus} which means interwoven. This implies that components are interdependent. Thus, the key in complex systems research is that there is a strong focus on \emph{interactions}~\citep{Gershenson:2011e}. Interactions in complex systems co-determine the future of systems, and thus limit predictability and the experimental testing of equations. It is not enough to know initial and boundary conditions, as interactions generate novel information and complex systems are not isolated: there are relevant changes from the outside and from the inside of the system. 

Complexity scales with number of elements, with number of interactions, with complexity of elements and with complexity of interactions~\citep[p. 13]{GershensonDCSOS}. For instance, the interaction between two people can be more complex than interactions between several people in a crowd. One of the challenges of complexity is to find proper trade offs, for example to reveal the optimal size of groups for a specific purpose. 

We are able to study complex systems because of computers and statistics. Only now we are able to build models which can take into account dozens or millions of variables and interactions. Considering large multidimensional spaces, it has become clear that simple rules can lead to complex dynamics~\citep{Wolfram:2002}.

The contributions that complexity has made in line with CSR include network theory~\citep{Csermely:2006,Newman:2010,Motter2012Networks}, statistical mechanics~\citep{stanley1987introduction}, agent based modelling~\citep{Bonabeau2002ABM}, and evolutionary dynamics~\citep{nowak2006evolutionary}.

There have been applications to most fields, including systems biology~\citep{Kauffman1993,Kitano2002SysBiol}, computer science~\citep{Berners-Lee1992World-Wide-Web,BrinPage1998}, economics~\citep{Arthur:2011}, social systems~\citep{EpsteinAxtell1996}, ecology~\citep{Ulanowicz1977}, and chemistry~\citep{Lehn:1990}.

The availability of ``big data"~\citep{Lynch2008Big-data:-How-d} is enabling us to contrast different models, so many of them can be rejected. Many biological and social theories were impossible to test because of lack of data. Now we are having not only the data, but the methods to analyze it. This is not about making ``soft" sciences harder, but about making them empirical. 

The pervasiveness of complex systems and the resulting technology is changing society. For example, mobile devices leave digital trails which can be exploited for different purposes~\citep{D4D2013}, including the verification of social theories. However, privacy concerns are yet to be resolved~\citep{Montjoye2013Unique-in-the-C}. We want progress, but at what cost?

\section{The Future}

The future is ripe with challenges. The XXIst century has been and will be a century of crisis, fast changes, urban and economic problems, limits of growth, instability, overpopulation, climate change, and several other challenges. Globalization is leading systems to become more and more interdependent. There are many problems that must be solved. To what extent CSR will contribute to the solution of these problems? To what extent CSR would be acknowledged, given the fact that it has already permeated into mainstream science? In practice, it does not matter. It is clear, however, that cybernetic and systemic concepts are necessary to solve future challenges~\citep{Helbing2012FuturICT:-Parti,Helbing2013Globally-networ}.

For understanding phenomena, we have to refine our descriptions in order to relate different scales (levels of abstraction). From an evolutionary perspective, we also need to develop a better understanding of transitions~\citep{Turchin1977,szathmary97,Scheffer2009Early-warning-s}, e.g. what makes the non-living to become living (from chemistry to biology), what makes the living creatures become conscious (from living systems to human consciousness), and what is the nature of the human spirit (as the highest level of consciousness). These are three main emergences in the big history of the universe.

There is also the need to build a closer relationship between natural sciences and the humanities. Ethics, esthetics, and other branches of philosophy, especially when they apply naturalistic approaches, are already successfully using notions of systems thinking. But if we consider the modern cognitive science and epistemology, it’s rather strange why they do not deal with human spirit---\emph{menschlicher Geist}~\citep{Knyazeva2009Nonlinear-Cobwe}. Why is the concept of human spirit lacking in the modern research? If we do not study it from a scientific point of view, other people will treat it from mysterious, esoteric, religious, and similar non-scientific perspectives.

A common language and a common vision are required. CSR has the potential of offering this to both sciences and humanities. Global problems require a combination of phenomenology and theory, of reductionism and holism. But to achieve this, a common language is needed~\citep{Knyazeva2008Synergetics}.

A shift in education is also necessary. It is still unclear how the education of the future will be, but it is clear that current education is failing. CSR has the opportunity to contribute to this effort.

\section{Discussion}

We are recovering from extremes of reductionism in science. For example, biology is recovering from the reductionist use of the results of molecular biology~\citep{Csermely:2006}. Extreme reductionism focusses on a single isolated protein, adds another and studies the interaction. This is nonsense because in real cells there are hundreds of other proteins interacting with both proteins and affecting their interactions. The extreme of focussing on isolated components of complex systems should be avoided. But also the other extreme should be avoided: the extreme of focussing on systems and forgetting about the data. In other words, our descriptions will be insufficient if we focus at a single scale, e.g. element or system. Focussing on multiple scales will give us a broader understanding of phenomena~\citep{BarYam2004}. This leads to three methodological comments:

\begin{enumerate}
\item We should be humble. The understanding of systems has limits. Let us focus on a protein in the brain, which interacts with other proteins. Imagine the brain is of a youth on a first date, excited, so also proteins are excited. The protein in the middle of the turmoil doesn't have a clue why, nor that it is in a brain, nor that the owner of the brain is on a first date, which is the cause of its current situation. If we think we are like a proteins, we start to understand how humble we should be when understanding systems at higher levels.

\item We have to be very cautious. There is a difference between finding solutions to problems and finding problems to solutions. Some models have been disproved with experimental data, but sometimes we stick to our models and we try to find the segment of the world which can be finally described by that model. 
As an example of these fallacies, Adam Smith’s invisible hand metaphor for market behavior may be cited, which became a basis of a large number of models in economy even in years when the complexity of market dynamics has already been well established.

\item We have to be open minded. For example, mathematicians, physicists and biologists can have different a understanding of phase transitions. But they are just using the same word for different descriptions. All of them are right and we should be aware of it. The same phenomenon can be described from a variety of contexts~\citep{Gershenson2002ua}, which is exemplified well by the Indian parable of the six blind men and the elephant. It will be more productive to be inclusive and consider different perspectives rather than being exclusive and reject all but one.

\end{enumerate}

From an evolutionary perspective, in science the best ideas are those that change society and endure. 
Metaphors can be used for providing novel descriptions. But in order to change systems we have to understand them. We need to ask good questions, and then to listen very carefully what nature replies. This is how all of science should be done.

\section{Conclusions}

CSR have strongly influenced all scientific disciplines. As an example, the term ``system" is used commonly in daily language. One of the breakthroughs of CSR involves the attempt to find commonalities across disciplines. Even when this was achieved to a certain degree, there is still a lack of a common language to communicate successfully, especially between the natural and social sciences.

Currently, the scientific study of complex systems has several commonalities with CSR. It could be argued that complexity has inherited many of the aims of CSR, and they can be distinguished roughly by complexity being dominated more by natural sciences and CSR more by social sciences, although there is a strong overlap. One of the aspects that has propagated complexity has been its ability to contrast its theories and dispose those that do not match observations. This is a challenge for CSR, where theories should also be contrasted with real data. Nevertheless, this is becoming feasible due to the increased accessibility to several sources of information and methods for analysing this data.

It is suggested that CSR researchers should be humble (since our knowledge and cognitive abilities are limited), cautious (not to believe blindly in our models), and open minded (towards other disciplines and approaches). As our future unfolds, CSR has the opportunity to solve relevant problems of our globalized society~\citep{laszlo2003systems}. This makes CSR an ambitious endeavor. However, in order to find our limits we have to go beyond them.

\section*{Acknowledgments}

We should like to\ thank the organizers of the 2012 European Meeting on Cybernetics and Systems Research for making the round table that led to this paper possible. Ray Ison provided useful comments.
C. G. was partially supported by SNI membership 47907 of CONACyT, Mexico. Work in P.C’s laboratory was supported by a research grant from the Hungarian National Science Foundation (OTKA-K83314).

\bibliographystyle{cgg}
\bibliography{carlos,sos,complex,COG,information,evolution,RBN,orgs}

\end{document}